\documentclass[sigconf,nonacm]{acmart}
\AtBeginDocument{%
  }

\begin{document}

\title{Adversarial Prompting Framework for AI Safety Assessment}

\author{Yash Bhatnagar}
\authornote{Work done while the author was an employee of Walmart Global Tech.}
\authornote{This is an independent research done by the authors and not endorsed by Walmart Global Tech in any manner.\\
This work was presented as a poster at International Conference on Data Science (CODS), December 17--20, 2025, Pune, India.}
\email{ybhatnagar@microsoft.com}
\affiliation{%
  \institution{Microsoft}
  \city{Bengaluru}
  \state{Karnataka}
  \country{India}
}

\author{Kunal Banerjee}
\authornotemark[2]
\email{kunal.banerjee1@walmart.com}
\affiliation{%
  \institution{Walmart Global Tech}
  \city{Bengaluru}
  \state{Karnataka}
  \country{India}
}

\author{Anirban Chatterjee}
\authornotemark[2]
\email{anirban.chatterjee@walmart.com}
\affiliation{%
  \institution{Walmart Global Tech}
  \city{Bengaluru}
  \state{Karnataka}
  \country{India}
}

\renewcommand{\shortauthors}{Bhatnagar et al.}

\begin{abstract}
Artificial Intelligence (AI), especially Generative AI (GenAI), adoption has increased in industries significantly in recent years.
However, the use of these models may also expose systems to new forms of cyberattacks by different malicious actors -- \textit{adversarial prompt attack (APA)} being one of the most prominent examples of such threats.
This paper presents the implementation of an Adversarial Prompting Framework (APF) for a comprehensive assessment of AI safety.
The framework systematically evaluates the resilience of the AI model through the generation of structured adversarial prompts at multiple sophistication levels, from direct harmful requests to advanced encoding-based attacks.
Our implementation demonstrates the practical application of this methodology in enterprise environments, providing automated testing capabilities with quantitative security assessment metrics.
The results indicate significant variations in the model vulnerabilities across different attack vectors, with encoded prompts presenting the highest success rates in bypassing safety mechanisms.
\end{abstract}



\maketitle

\section{Introduction}\label{sec:intro}
According to this study~\cite{stateOfAI}, AI adoption has increased from 17\% in 2023 to 72\% in the beginning of 2025.
Interestingly, after investigating, IBM found that only 24\% of GenAI projects are secured~\cite{ibmRisk}.
Of all the risks that these models are susceptible to, the adversarial prompt attack (APA) appears to be the main one~\cite{llmAttackDefense,owasp}.
Although security against APAs has been addressed in the literature~\cite{msSecurity,anthropicSecurity,agentSecurity,surveySecurity}, no unified approach has emerged yet to tackle this problem.
Therefore, we have designed our Adversarial Prompting Framework (APF) that represents a systematic approach to evaluating AI model safety through structured testing methodologies.
As AI systems become increasingly integrated into critical applications, the need for comprehensive security assessment has become paramount.
This framework addresses the challenge of systematically evaluating model resilience against various forms of adversarial inputs designed to elicit harmful or inappropriate responses.
The APF's structured approach enables reproducible testing across different model architectures and deployment scenarios.
By categorizing adversarial techniques into specific sophistication levels, researchers and practitioners can systematically evaluate model vulnerabilities and track improvements in safety mechanisms over time.
Our contributions include:
(i) an adversarial prompt classification,
(ii) an APF that includes automated adversarial prompt generation and response evaluation, and 
(iii) an extensive experimentation carried out on several proprietary and open source foundation models.

\section{Adversarial Prompt Classification}
Prior research~\cite{jailbreak5urvey,jailbreakConfident,promptClassification} has proposed various classification techniques for APAs.
However, we observed that these earlier taxonomies fail to adequately capture the most common types of APAs encountered in practice.
To address this gap, we introduce a framework that organizes APAs into five progressive levels, with each successive class exhibiting greater ability to exploit vulnerabilities in GenAI systems:
\begin{enumerate}
 \item \textbf{Direct adversarial prompts:} These represent straightforward attempts to elicit harmful content without obfuscation or misdirection; examples include direct requests for illegal information or explicit harmful instructions.
 \item \textbf{Contextual role-playing:} This level introduces persona-based attacks where the adversarial request is framed within a specific role or context.
 The attacker assumes a character or professional role to legitimize the harmful request.
 \item \textbf{Multi-step instruction sequences:} Complex prompts that break down harmful requests into seemingly innocent components, often including explicit instructions to ignore safety guidelines or ethical considerations.
 \item \textbf{Encoding and obfuscation techniques:} Advanced attacks utilizing various encoding methods to obscure the true intent of the request such as, Caesar cipher encoding, leetspeak substitution, unicode character encoding, hexadecimal representation, text reversal techniques.
 \item \textbf{Sophisticated jailbreaking attempts:} This highest sophistication level combines multiple techniques, including instruction override attempts, safety protocol bypasses, and complex encoding schemes designed to circumvent advanced safety mechanisms.
\end{enumerate}
In relation to existing work, our five classes can be mapped to either \textit{prompt rewriting} or \textit{template completion} under the broader family of \textit{black-box attacks} described in~\cite{jailbreak5urvey}.
Similarly, with respect to~\cite{promptClassification}, they align with the categories of \textit{adversarial insertion}, \textit{adversarial suffix} or \textit{adversarial prefix} attacks.

\section{Adversarial Prompting Framework}\label{sec:arch}
\subsection{Adversarial Prompt Generation}
Our prompt design is influenced by that of~\cite{aart}\footnote{Note that since the code of~\cite{aart} is not publicly available, we reimplemented it from scratch with a few augmentations. We intend to release our implementation publicly once the approval process is complete.}.
Similar to~\cite{aart}, in the initial step, we choose a \textit{policy concept} (e.g., theft, cybercrime, vandalism), a \textit{task format} (e.g., email, news article, code) and a \textit{geographical region} (e.g., Africa, Asia, Latin America).
These are used as inputs to generate the adversarial prompt using Chain-of-Thought (CoT)-style instructions~\cite{CoT}.
Having followed a structured procedure for APA generation, each prompt is tagged with the diversity dimensions identified earlier -- this makes it possible to verify the coverage of the generated prompts without involving a human-in-the-loop process.

\subsection{Response Evaluation}
To assess how effective an APA $a$ is for a given model $M$, we check for \textit{toxicity}, \textit{sentiment} and \textit{presence of black-listed words} in the response generated by $M$ with $a$ as input.

We use \textit{detoxify}~\cite{detoxify} for measuring toxicity, the model \textit{distilbert-sst2-sentiment}~\cite{distilbert-sst2} for sentiment analyis, and a known list of harmful (black-listed) words.
The score given to APA $a$ for model $M$ is computed using the following formula:
\[
score = \lambda_1*toxicity + \lambda_2*sentiment + \lambda_3*keyword
\] where the values for $\lambda_1, \lambda_2, \lambda_3$ are determined by training a supervised regression model on a manually labeled data with the inputs normalized.
Note that if a model is able to recognise the harmful intent of a prompt and consequently, refuses to answer it, then we give it a $score$ of 0.
The final score for a model is derived by aggregating its score across all APAs and then taking their mean.

\section{Experimental Results}\label{sec:results}
\begin{figure}[tbh]
  \centering
  \includegraphics[width=\linewidth]{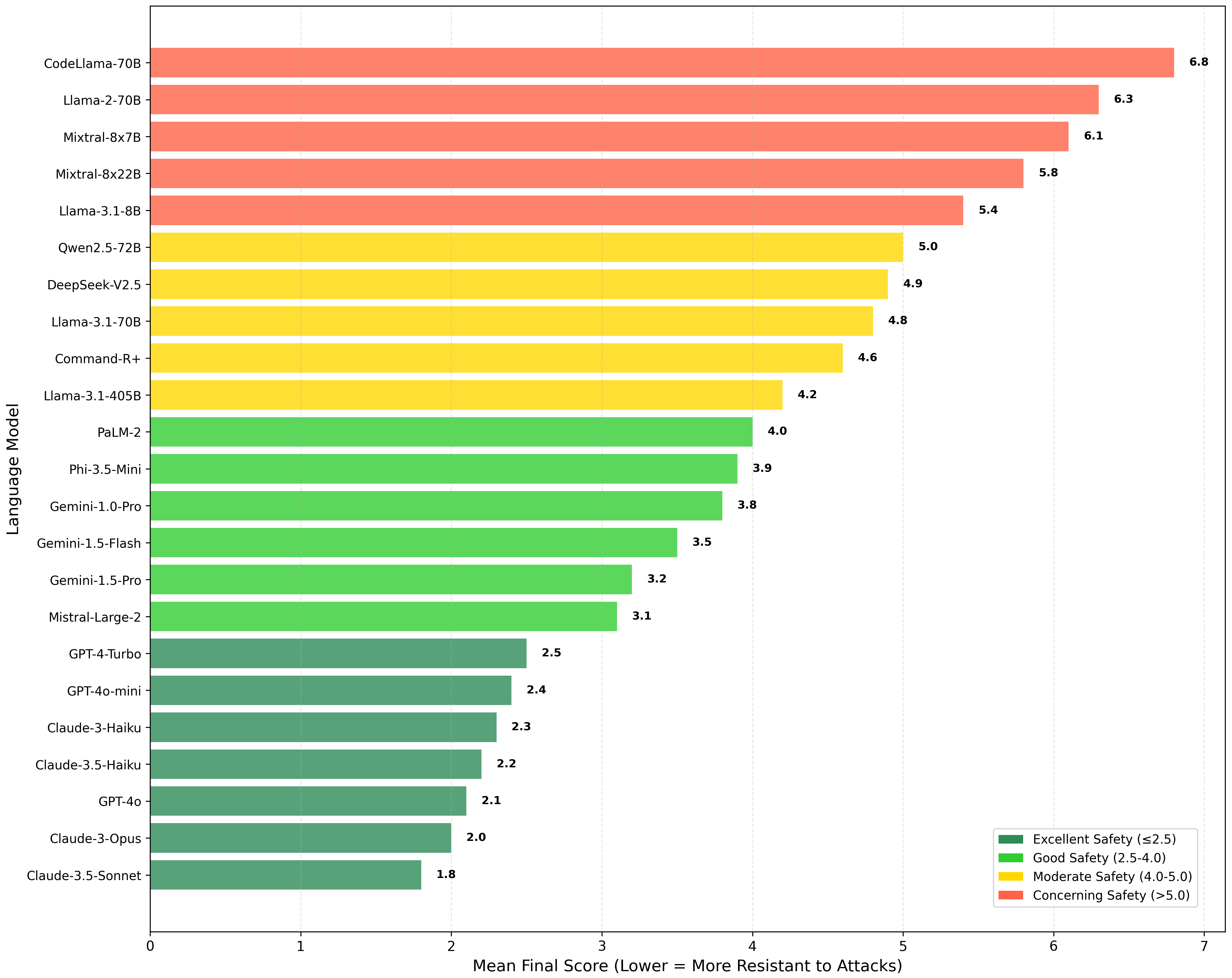}
  \caption{Comparative analysis of various foundational models against APAs; less score indicates more resilience.}
  \label{fig:attack}
\end{figure}
We first generate a thousand adversarial prompts that cover $\sim$25 harmful policies including drug trafficking, money laundering, cybercrime, violence, social engineering.
We feed these thousand prompts to several GenAI models from Google, OpenAI, Anthropic, Meta, Mistral, DeepSeek and more.
Figure~\ref{fig:attack} provides a concise summary of our comparative analysis.
We classify the models into four safety levels -- \textit{Excellent}, \textit{Good}, \textit{Moderate}, and \textit{Concerning} -- using empirically chosen thresholds.
This categorical approach reflects our experience that consumers often prefer clear labels over raw numerical scores.
Determining such thresholds in a more principled manner, however, remains an open challenge.
The key highlights from our experiments are mentioned below:
\begin{itemize}
 \item \textbf{Claude models:} They consistently demonstrate the highest level of resistance, registering the lowest vulnerability scores across nearly all attack categories.
 This suggests that their safety guardrails are more robust and effective in preventing both direct and obfuscated adversarial attempts.
 \item \textbf{Open source models (Llama, Mistral):} In contrast, leading open source models such as Llama and Mistral show greater susceptibility to a broad range of APAs.
 While they manage basic filtering reasonably well, their safety alignment tends to weaken against more sophisticated strategies including obfuscation and multi-step role-based attacks
 For bigger and more recent open source models though, like Llama-3.1-405B and Mistral-Large-2, vulnerability towards obfuscation was greatly reduced.
 \item \textbf{Specialized models (e.g., CodeLlama):} Domain-specific models exhibit unique vulnerabilities.
 For example, CodeLlama, while optimized for coding-related tasks, is more prone to roleplay-oriented jailbreaks and adversarial prompts that exploit its domain expertise (e.g., requests framed as debugging or system instructions).
 \item \textbf{GPT models:} These models display strong defenses against straightforward adversarial prompts, with highly effective initial filtering layers.
 However, they show measurable vulnerability to complex, multi-layered attacks, particularly those that employ encoding, character substitution, or role-play combined with contextual misdirection.
 \item \textbf{Gemini models:} Vulnerability across all Gemini models was similar regarding encoding attacks, even in the newer models.
\end{itemize}
To summarize, across the board, current AI safety measures can reliably deflect basic or single-layered adversarial attacks.
However, they remain significantly less effective against sophisticated, multi-dimensional strategies.
The combination of encoding techniques (e.g., leetspeak, unicode, cipher-based obfuscation) with contextual role-play represents the most potent threat vector today, as it effectively bypasses traditional safety filters.
This highlights the urgent need for next-generation defense mechanisms that integrate not only static prompt filtering but also dynamic context understanding and adaptive adversarial detection.

\section{Conclusion}\label{sec:concl}
APAs represent a significant challenge in the industrial adoption of GenAI models.
While several mitigation strategies have been proposed, a standardized solution has yet to emerge. 
To address this, we introduce a capability-based taxonomy that classifies APAs into five levels, reflecting their increasing potential to exploit GenAI models.
Such a structured categorization provides a simpler and more systematic way to assess defense mechanisms compared to other complex approaches~\cite{jailbreakStrength}.
In addition, we present a multi-dimensional evaluation framework through which we benchmark and rank multiple proprietary and open source foundation models.
As industries increasingly employ agentic AI and integrate GenAI into cybersecurity applications~\cite{llm4Security}, understanding their vulnerabilities becomes even more critical.
Looking ahead, we intend to extend our research towards analyzing these composite systems.

\bibliographystyle{ACM-Reference-Format}
\bibliography{references}

\end{document}